# Enhanced Relative Cooling Power and large inverse magnetocaloric effect of Cobalt ferrite nanoparticles synthesized by auto-combustion method

Youness Hadouch[1, 2, *], Daoud Mezzane[1, 2], M'barek Amjoud[1], Lahoucine Hajji[1], Yaovi Gagou[2], Zdravko Kutnjak[3], Valentin Laguta[4, 5], Yakov Kopelevich[6], Mimoun El Marssi[2].

*1 Laboratory of Innovative Materials, Energy and Sustainable Development (IMED), Cadi-Ayyad University, Faculty of Sciences and Technology, BP 549, Marrakech, Morocco.*
*2 Laboratory of Physics of Condensed Matter (LPMC), University of Picardie Jules Verne, Scientific Pole, 33 rue Saint-Leu, 80039 Amiens Cedex 1, France.*
*3 Jozef Stefan Institute, Jamova Cesta 39, 1000 Ljubljana, Slovenia.*
*4 Institute of Physics AS CR, Cukrovarnicka 10, 162 53 Prague, Czech Republic.*
*5 Institute for Problems of Materials Science, National Ac. of Science, Krjijanovskogo 3, Kyiv 03142, Ukraine.*
*6 Universidade Estadual de Campinas-UNICAMP, Instituto de Física "GlebWataghin", R. Sergio Buarque de Holanda 777, 13083-859 Campinas, Brazil*

*Corresponding author:

E-mail: hadouch.younes@gmail.com; youness.hadouch@etud.u-picardie.fr

ORCID: https://orcid.org/0000-0002-8087-9494

Tel: +212-6 49 97 06 74

***Highlights:***

- Cobalt ferrite nanoparticles were elaborated by the sol-gel auto combustion method;
- Large magnetocaloric temperature change is observed in CFO nanoparticles;
- High Relative Cooling Power (RCP) coefficient and the temperature-averaged entropy change (TEC) are obtained in CFO.

***Abstract:***

This work focuses on the microstructure, magnetic properties and magnetocaloric effect of $CoFe_2O_4$ (CFO) nanoparticles elaborated by the sol-gel auto combustion method. The XRD investigation indicates that CFO is crystallized in a cubic spinel structure with the $Fd\bar{3}m$ space group and the SEM micrograph shows fine quasi-spherical nanoparticles with an average grain size of 160 nm. The temperature dependence of the Raman spectra reveals the ferromagnetic to paramagnetic conversion started from 723 K. The magnetization temperature dependence reveals the Curie temperature at $T_C$= 785 K. Large value of magnetocaloric temperature change of ΔT =11.2 K with a high RCP of 687.56 J Kg$^{-1}$ were measured indirectly via the Maxwell approach making our CFO nanopowder suitable candidate for both environmentally friendly magnetic refrigeration and medical applications at ambient temperature.

*Keywords:* Cobalt ferrite; sol-gel self-combustion; Nanoparticles; First order; Magnetocaloric effect; Relative Cooling Power.

*Acknowledgements:*

The authors gratefully acknowledge the generous financial support of CNRST Priority Program PPR 15/2015, the European Union's Horizon 2020 research and FAPESP and CNPq, Brazilian agencies.

*Formatting of funding sources:*

CNRST Priority Program PPR 15/2015;

The European Union's Horizon 2020 research;

*Introduction:*

Recently, the magnetocaloric effect (MCE) has attracted much interest from researchers for its use in various industrial applications. It was discovered by German physicist Emil Warburg in 1881[1]. Scientists are now devoting many efforts to ecological and friendly magnetic materials used in heating and cooling technologies[2], [3]. The MCE is an intrinsic property of magnetocaloric materials in which a reversible and adiabatic temperature change is caused when they are exposed to a varying magnetic field[4]. It can be evaluated as an isothermal entropy change $\Delta S$ and/or as an adiabatic temperature change $\Delta T$[4].

Several classes of magnetic materials displayed a magnetocaloric effect in the literature, including perovskite manganites, manganite alloys, and spinel ferrites. The magnetocaloric materials can be classified according to the order of the phase transition; first- (FOPT) or second-order (SOPT) type phase transitions. Usually, the FOPT materials exhibit large magnetic entropy change and adiabatic temperature change[5]. Trung et al. investigated the effect of phase transition order on $\Delta S$ and $\Delta T$ for $Mn_{1-x}Cr_xCoGe$ alloys and demonstrated that the first order enhances the magnetocaloric effect[6].

Recently, the perovskite manganite $La_{0.5}Ca_{0.5-x}\square_xMnO_3$(LCMO) exhibiting an interesting MCE, with a $\Delta T$ of 5.6 K, is reported [7]. Meanwhile, several works have shown that spinel ferrites with a second-order magnetic transition are effective instruments for magnetic refrigeration (MR) applications. Oumezzine et al. observed a significant magnetic entropy change $\Delta S_M$ of 1.61 J kg$^{-1}$ K$^{-1}$ at 50 kOe in $Zn_{0.6}Cu_{0.4}Fe_2O_4$ with a RCP of 289 J kg$^{-1}$[8].

Moreover, $Ni_{0.7}Zn_{0.3}Fe_2O_4$ showed a moderate ($-\Delta S_M$) of 1.39 J kg$^{-1}$ K$^{-1}$ at 25 kOe with a high RCP value of 68 J kg$^{-1}$ as reported by Anwar et al. [9].

Among spinel ferrites, cobalt ferrite ($CoFe_2O_4$) has gained great scientific interest owing its moderate saturation magnetization, high coercivity, good magnetostrictive properties, high Curie temperature ($T_C$=520 °C), electrical insulation with low eddy current loss, and chemical stability [10]–[12]. Benefiting from these properties, $CoFe_2O_4$ (CFO) find a variety of applications in several domains such as data storage, sensors, microwave devices, high-frequency applications, catalysis, magnetic refrigeration and biomedical field [10], [13]–[16]. More particularly, in refrigeration applications, CFO intended to replace carbon technology to limit global warming and reduce greenhouse gas emissions[17], [18]. CFO nanoparticles are used in cancer therapy by hyperthermia (cancer treatment) for biomedical applications. CFO nanoparticles were inserted into tumor tissue and subsequently heated after being exposed to an external alternating magnetic field [19], [20].

Crystallographically, CFO has an inverse spinel structure where octahedral sites are occupied by $Co^{2+}$ cations, while $Fe^{3+}$ cations are equally shared between tetrahedral and octahedral sites[21]. It is known that the magnetic properties are sensitive to the distribution of these cations, which is influenced by the preparation method of CFO nanoparticles [22], [23]. Furthermore, ferrite properties depend greatly on grain size and microstructure, which can also be monitored by the synthesis route [24], [25]. For example, the saturation of magnetization ($M_s$) of $NiFe_2O_4$ was increased with increasing ferrite grain size, as reported by Li et al. [25]. Furthermore, several studies reported that the particle size tuned the MCE of magnetic nanoparticles [26], [27]. For example, Yin et al. reported that the $\Delta S_M$ in $HoCrO_3$ decreases ($\Delta S_M$ =8.73, 7.22, 7.77, and 6.70 J kg$^{-1}$ K$^{-1}$) and the refrigerant capacity (RC = 388, 354, 330, and 310 J kg$^{-1}$) for the 60 nm, 190 nm, 320 nm, and 425 nm size particles, respectively. Accordingly, the synthesis methods are crucial in controlling the magnetic properties of ferrites powders and their applications. CFO is typically elaborated using various methods, including co-precipitation sol-gel, ceramic techniques and auto combustion utilizing different fuels [28]–[30]. Researchers have recently focused on coupling sol gel and auto combustion method, as it produced ferrite samples with high chemical homogeneity, good purity, and high crystallization. In addition, it demands basic equipment, a simple preparation process, a short processing time and low external energy consumption [31], [32].

It is worth noting that the magnetic characteristics of most CFO have been examined over a wide range of temperatures, from room temperature to as low as 4 K. However, the majority of ferrite's applications are suitable at room temperature or above, for example, medical application in which magnetocaloric properties can induce local hyperthermia at cancer sites [33]. Note that very few reports describe the MCE properties of cobalt ferrite at high temperatures. Herein, we present an investigation at the high-temperature range from 300 K to 900 K on structural, magnetic and magnetocaloric properties of $CoFe_2O_4$ elaborated by combining sol-gel auto combustion using ammonia as a neutralizer agent, ethylene glycol as polymerizing agent and acetic acid as fuel.

*Chemical synthesis:*

$CoFe_2O_4$ nanoparticles were prepared using the sol-gel self-combustion method: stoichiometric amounts of Cobalt Nitrate $Co(NO_3)_3.6H_2O$ and Ferric Nitrate $Fe(NO_3)_3.9H_2O$ were dissolved in 2-Methoxyethonal under constant magnetic stirring for 30 mn. Then ammonia was slowly dropped to the solution to adjust the *pH* to 10. A 1:1 molar ratio of acetic acid (fuel agent) and ethylene glycol (polymerizing agent) mixture was separately prepared and added to the solution. After four hours of continuous stirring at 90°C, the brown solution was completely turned into a dark, puffy, porous gel. Then, the gel was transferred to a furnace, and by heating at 120 °C, it simultaneously burnt in self-propagating combustion until it was completely transformed into fine powder. Finally, the powder was annealed at 700 °C for two hours using a controlled heating rate of 2 °C/min under an air atmosphere.

*Materials and methods:*

The XRD patterns of CFO ceramic were obtained by X-ray diffraction using the Panalytical X-Pert Pro with Cu-Kα radiation ($\lambda = 1.54059$ Å) at room temperature. The grain morphology of the ceramic was examined using a scanning electron microscope (SEM, Philips XL30) having an attachment for energy dispersive X-ray spectroscopy (EDS). The Raman spectra were recorded using a micro-Raman Renishaw spectrometer equipped with a CCD detector. Magnetic properties were performed using a Physical Property Measurement System (PPMS-DynaCool) Quantum Design apparatus in the 300-900 K temperature range under a magnetic field range of 0-25 kOe. The magnetization was measured using the vibrating sample magnetometer (VSM) method integrated into system. The magnetocaloric study was carried out by the indirect method using the recorded *M–H* hysteresis loops.

*Results and discussion:*

*Morphological and structural analysis:*

The SEM image of the as-prepared CoFe$_2$O$_4$ nanoparticles shows agglomerates of regular near-spherical particles with average grain sizes of 160 nm, estimated by ImageJ® software as shown in Fig.1. The large particle sizes are due to the high heating temperatures released during the combustion process[11]. It should be noted that the magnetic properties of cobalt ferrites are very sensitive to grain sizes.

EDX analysis was performed to verify the elemental composition, and the distribution of elements is shown in Fig.2. The presence of metal elements Co, Fe and O is revealed by qualitative analysis (the presence of carbon is due to the metallization process). The atomic and weight percentages of the elements present in the sample are shown in the inset, and the obtained atomic percentages agree well with the expected ratios. The oxygen content is found to be 3.9683 atoms per cell.

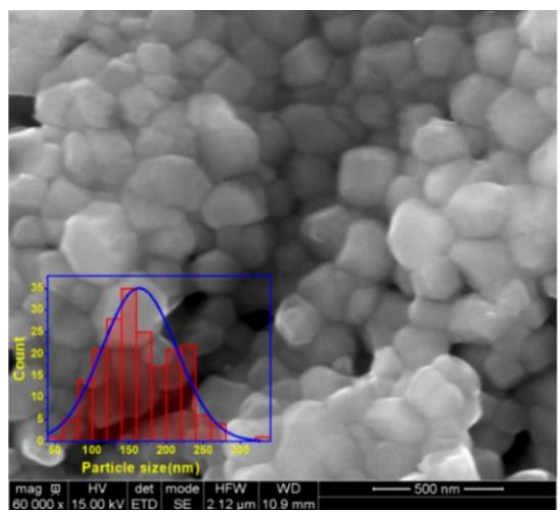

Fig. 1 SEM micrograph of CFO nanoparticles, the inset to the image shows the particle size distribution

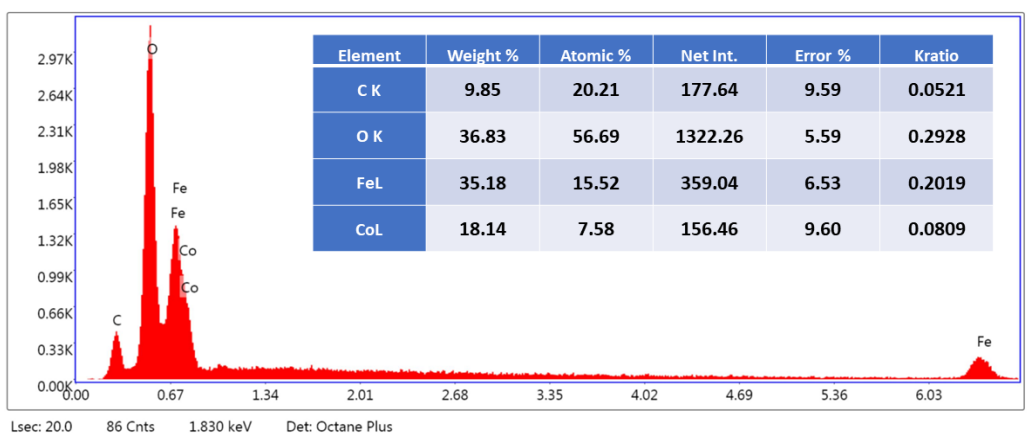

Fig. 2 EDS spectrum of CFO nanoparticles.

Fig. 3 shows the room temperature XRD spectra of the CFO powder calcined at 700 °C for 2h. The obtained diffraction peaks are well-defined proving that our sample is well crystallized. All the diffraction peaks can be indexed to the cubic crystal structure of cobalt ferrite (JCPDS card no. 22-1086)[34], and no peak related to impurities was detected. The experimental XRD pattern of CFO was refined using the Rietveld refinement program via FullProf software and indexed by a cubic spinel structure with the space group ($Fd\bar{3}m$), and the output data proved an excellent fit between the calculated and observed diffractograms. The "a" lattice parameter of CFO is found to be 8.4013 Å. According to Scherrer's equation[14], [35], [36], the average crystallite size (D) was estimated from the full width at half-maximum of the conspicuous (311) reflection and found to be 57.6 nm and the grain size of 160 nm determined using ImageJ® software. Table 1 lists the refined structural parameters of CFO powder.

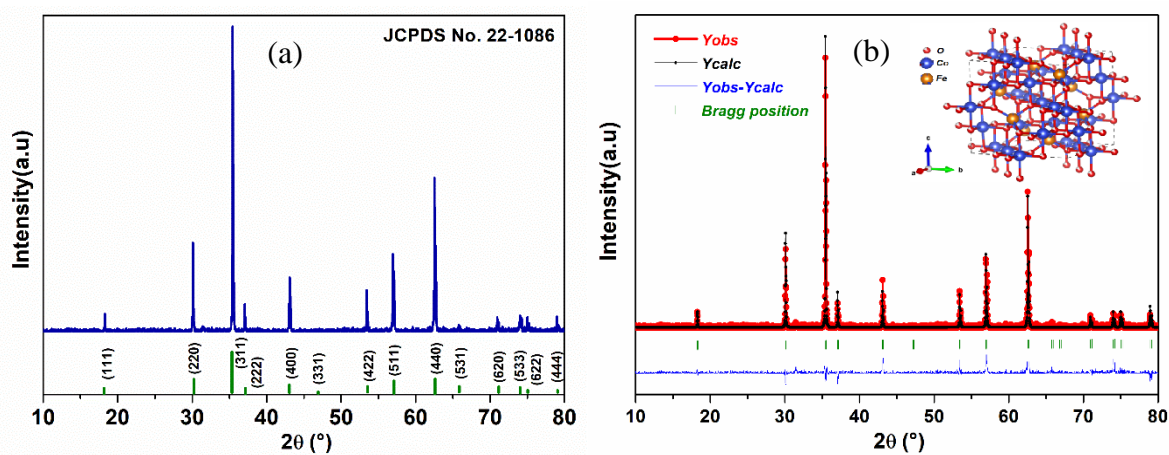

Fig. 3 **a** X-ray diffraction patterns of CFO powder; **b** Rietveld fitted X-ray diffraction patterns of CFO nanoparticles

Table 1 Refined structural parameters for $CoFe_2O_{4-\delta}$ nanoparticles at room temperature

| Structure Spinel | Unit cell parameters | | | Atom fraction | | | Molecular formula F | Average crystallite size(nm) By XRD | Average grain size(nm) by SEM |
|---|---|---|---|---|---|---|---|---|---|
| | a(Å) | Angle (°) | V(Å³) | Co | Fe | O | | | |
| $Fd\bar{3}m$ | 8.4013 | $\alpha = \beta = \gamma = 90$ | 592.9792 | 0.1446 | 0.2892 | 0.5661 | $CoFe_2O_{3.9627}$ | 57.6 | 160 |

Raman spectra were collected in the temperature range of 303-823 K to study the temperature effect on the spinel structure of the CFO, as shown in Fig.3. The group theory analysis predicts the following ten optical phonons for the spinel structure: $5T_{1u} + A_{1g} + E_g + 3T_{2g}$. The five $T_{1u}$ modesare IR active, while the other modes ($A_{1g} + E_g + 3T_{2g}$) are Raman active assigned to the motion of oxygen ions, A-site and B-site in the spinel structure[37]. Moreover, the $A_{1g}$ mode is assigned to symmetric stretching of the oxygen anions along the Fe-O (Co-O) in the tetrahedral sub-lattice. $E_g$ and first $T_{2g}(1)$ modes belong to the symmetric and asymmetric bending of the oxygen anions. The second $T_{2g}(2)$ mode arise due to asymmetric stretching of the Fe-O (Co-O) bonds, and the third $T_{2g}(3)$ corresponds to the translational shift of the entire $FeO_4$ tetrahedron.

At room temperature, $A_{1g}(1)$ and $A_{1g}(2)$ modes located at frequencies 616 and 693 cm$^{-1}$ are assigned to the stretching vibrations of the Fe–O and Co–O bonds in tetrahedral sites, respectively. At the frequencies below 600 cm$^{-1}$, the Raman modes (~208, ~303, ~468, and ~568 cm$^{-1}$) correspond to $E_g$ and $T_{2g}$ revealing the vibration of the spinel structure[38].

By heating, the $A_{1g}(1)$ and $A_{1g}(2)$ become softened and broad, resulting in the thermal effect on cation distribution in ferrite systems. In addition, the $A_{1g}(1)$ mode shifts to lower frequencies due to the migration of cations in a tetrahedral structure. At the same time, $E_g$ and $T_{2g}(3)$ phonon modes show the blue-shifting tendency associated with the cation migration in octahedral sites. In reality, the $Fe^{3+}$ cations move from O-site to T-site, while the $Co^{2+}$ cations move in the opposite direction from T-site to O-site. This dislocation of cations causes a disorder in the CFO structure that makes an inequality between these two valances.

This phenomenon increases with the increase of the temperature leading to disruption of the order of the cations. Thus, the magnetic properties of the CFO might be destroyed, giving rise to the paramagnetic phase at high temperatures [39], [40].

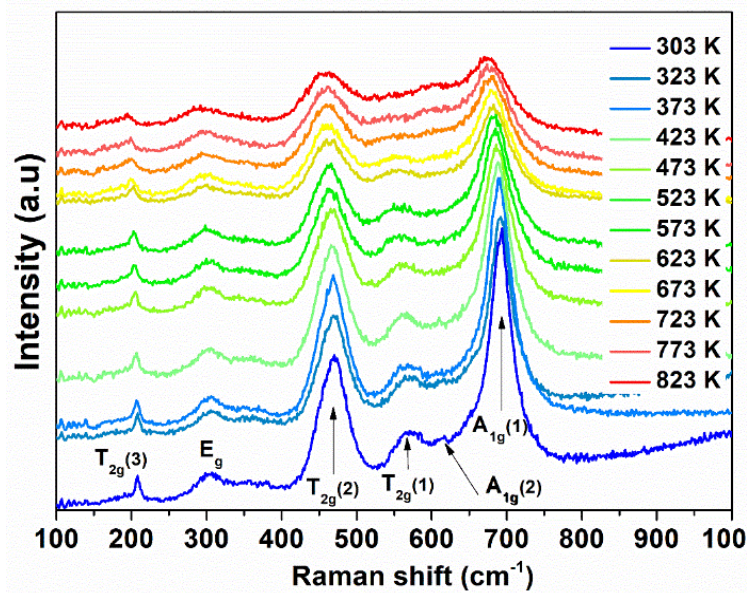

Fig. 4 Raman spectra of the CFO nanoparticles for several selected temperatures ranging from 303K to 823 K.

*Magnetic properties:*

*Temperature dependence of magnetic properties*

Fig. 4 (a) plots the temperature dependences of ZFC (blue curve) and FC (red curve) magnetization curves obtained in CFO powder under a magnetic field of 0.4 Oe. At 300 K, the magnetization was 8.16 emu g$^{-1}$, decreasing with increasing temperature. At around 800 K, the magnetization drops sharply to zero, which corresponds to the phase transition from ordered (FM) to disordered (PM) state as reported by Franco and Silva [33]. This discontinuous jump of magnetization reveals that our CFO powder exhibits a first order FM-PM phase transition at Curie temperature $T_C$. The Tc was determined from the derivative dM/dT(brown curve) and found to be 785 K Fig. 4 (a).

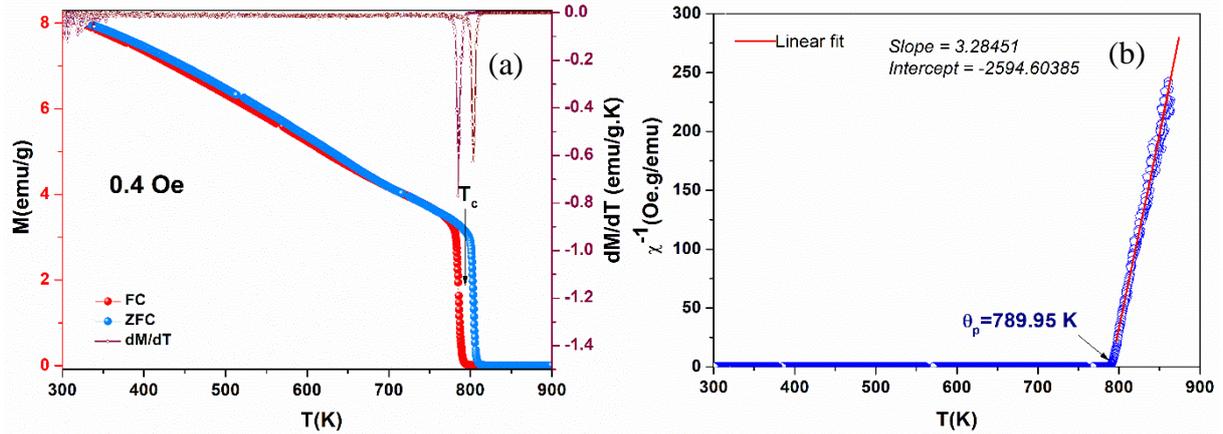

Fig. 5 The FC-ZFC curves (a) and reciprocal magnetic susceptibility $\chi^{-1}$ (T) (b), for the CFO sample measured at an applied magnetic field of 0.4 Oe.

The Tc value is confirmed using the Curie law (equation 1) by fitting the thermal variation of the inverse of the magnetic susceptibility ($\chi^{-1} = \frac{H}{M}$) plotted at 400 Oe:

$$\chi = \frac{C}{T - \theta_p} \ (Eq\ 1)$$

C denotes the Curie constant, while $\theta_p$ represents the Weiss temperature [41].

From the slope of the linear fit plot of $\chi(T)$ in the paramagnetic phase, the C is estimated at 0.30 K emu g$^{-1}$. The corresponding $\theta_p$ is 789.95 K that is closer to the Tc (785 K), as shown in Fig. 4 (b).

Fig. 5 displays the hysteresis loops (M–H) of CFO nanoparticles, measured under an applied magnetic field of 25 kOe and recorded at different temperatures (300 K - 900 K). It should be noted that the magnetic measurements were corrected to account for the demagnetization field and the internal field is calculated using the following expression [42]:

$$H = H_{app} - N.M$$

where $H_{app}$ denotes the externally magnetic field, M denotes magnetization, and N is the demagnetization tensor that depend on the geometry of particles.

In our case, a powder of randomly packed spherical particles could be a good approximation, and the demagnetization factor may be expressed as [43]:

$$N = \frac{1}{3} + f(D_z - \frac{1}{3})$$

where $D_z$ denotes the demagnetization factor of the powder sample's geometrical shape and f denotes the relative density. For spherical shape, the demagnetization factor is equal to 1/3[43]. The corrections have been made to the M versus H measurements for each temperature, and it appears that there is only a 0.0002 Oe difference between the internal H and external $H_{app}$ magnetic applied field.

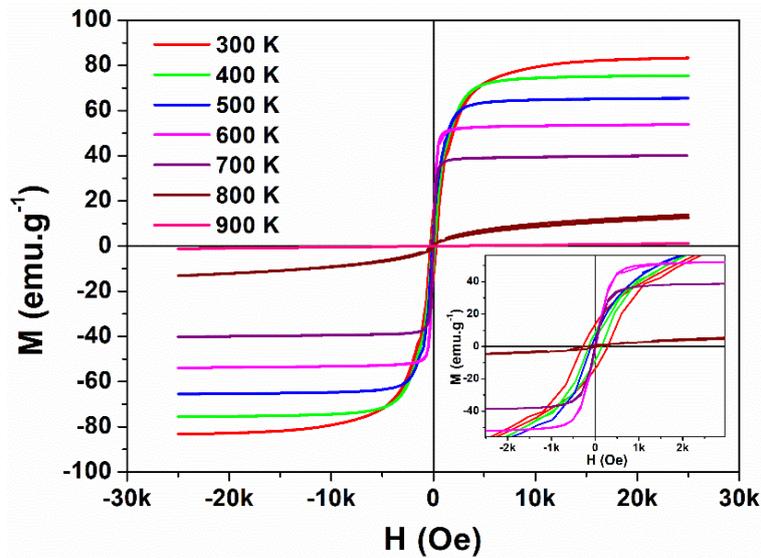

Fig .6 Isothermal magnetization curves of CFO at different temperatures under an applied magnetic field of 25 kOe

For T<Tc zone, a saturated hysteresis loop is observed, which describes CFO's ferro/ferrimagnetic character. The saturation of magnetization is attained even at 20 kOe. Hence, all the spins are oriented in parallel beyond this magnetic field value. At room temperature (300K), the CFO sample exhibits a saturated magnetization $M_s$ of 83 emu/g. Magnetization decreases as the temperature increases and disappears completely when T>Tc, indicating the appearance of paramagnetic character [7], [41]. This decrease is accompanied by a decrease in $M_s$, remnant magnetization ($M_r$) and coercive field ($H_c$), being equal to zero at close to $T_c$. Nevertheless, this is not a hint of superparamagnetism because the blocking temperature $T_b$ for CFO particles of 160 nm is much higher than $T_c$ ($T_b=KV/25k_B$, where K, V, and $k_B$ are anisotropy constants, particle volume, and Boltzmann constants, respectively) [33]. The details of the hysteresis curve for various temperatures are shown in the inset. The high $M_s$ (83 emu g$^{-1}$) observed at low $H_c$ (284 Oe) reveals the soft magnetic nature of CFO. This could be useful to design multifunctional devices to change the magnetization with a small external magnetic field [38]. The high obtained $M_s$ is due to the large particle sizes (160 nm) compared with CFO nanoparticles (35 nm) elaborated by hydrothermal that shows a $M_s$ of 68.5 emu g$^{-1}$ [11]. Indeed, the larger grains tend to consist of more magnetic domains. However, domain wall movement magnetization requires less energy than domain rotation magnetization. Therefore, the magnetization or demagnetization of the material is easy by the domain wall movement with a large grain size. As a result, a large grain sample is predicted to have low $H_c$ and high $M_s$ as reported by [25]. Subsequently, our sample shows improved results than those reported by other works using different synthesis methods of CFO, as summarized in table 2. Specifically, CFO elaborated by sol-gel auto combustion shows a $M_s$ of 69.59 emu g$^{-1}$ under a magnetic field of 10 kOe with a high $H_c$ of 647 Oe [44]. On the other hand, using solid-state method, Rajath Varma obtained a $M_s$ value of 82 emu g$^{-1}$ under a high applied field of 60 kOe [45]. As a result, the different values of $M_s$ and $H_c$ of all reported materials (Table2) demonstrate the synthesis method's effect or the microstructure of CFO on their magnetic properties.

Table 2 Comparison of magnetic properties Ms and Hc at 300K of CoFe$_2$O$_4$ particles obtained by various synthesis processes

| Synthesis method | Particle size (nm) | Magnetic field (kOe) | M$_s$ (emu g$^{-1}$) | H$_c$ (Oe) | References |
|---|---|---|---|---|---|
| Sol-gel auto combustion | 160 | 25 | 83 | 284 | This work |
| sol–gel auto combustion | 25 | 10 | 69.59 | 647 | [44] |
| Wet chemical route | 21 | 15 | 68 | 1250 | [28] |
| Co-precipitation | 34-40 | 15 | 74 | 650 | [11] |
| Hydrothermal | Rods (length: 139.63 / width: 35.23) | 15 | 68.5 | 1250 | [11] |
| Solid state | 290 | 60 | 82 | 304 | [45] |
| Sol–gel | 36.5 | 30 | 66.7 | 1163.9 | [14] |
| Polymer complex | 32 | 60 | 67 | 1625 | [45] |

*Magnetocaloric properties*

In order to investigate the magnetocaloric effect of our material, the indirect method based on M(H) measurement was performed. The magnetic entropy changes ΔS and the adiabatic temperature changes ΔT, versus the magnetic field (H), are given by the Maxwell Eqs. 2 and 3. Using the upper branches of all the recorded M(H) curves at each measurement temperature, the polynomial fits allow us to determine the variation of magnetization as a function of temperature ($\frac{\partial M}{\partial T}$) under various applied magnetic fields between 5 and 24 kOe. The resulting curves of ΔS and ΔT are plotted in fig. 7 (a) and (b), respectively.

$$\Delta S(T,H) = \int_0^H \left(\frac{\partial M}{\partial T}\right)_H dH \quad (Eq2)$$

$$\Delta T(T,H) = \int_{H1}^{H2} \frac{T}{C_P}\left(\frac{\partial M}{\partial T}\right)_H dH \quad (Eq3)$$

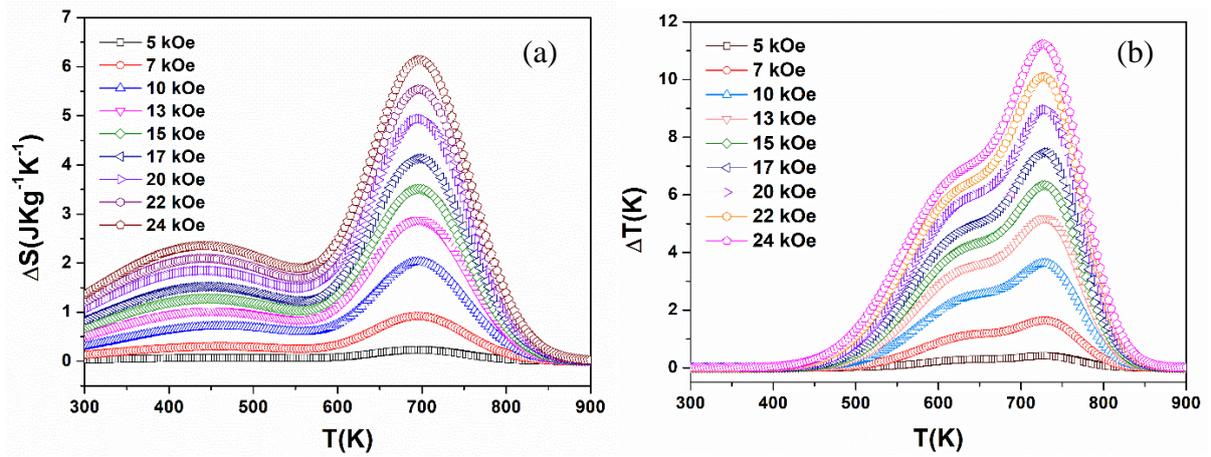

Fig. 7 Magnetic entropies change ΔS and the adiabatic temperature changes ΔT as a function of temperature under various magnetic fields for $CoFe_2O_4$ powder

Both curves reveal two magnetic anomalies. The anomaly that occurred at 700 K corresponds to the FM-PM phase transition. However, the first anomaly observed at 600 K on the ΔT curves and at 450 K on the ΔS curves can be attributed to lattice distortion and domain walls dynamics or pinning of domains. This process is activated by temperature, leading to the maximal reorientation at Curie temperature. Moreover, we can also observe on the M(H) measurement (Fig. 5) the vertical jump straight up of the magnetization M as shown on the ΔT reordered curve at 600 K. Maximal entropy of 6.2 $J.kg^{-1}K^{-1}$ is observed at the FM-PM phase transition temperature that corresponds to an adiabatic electrocaloric temperature change of 11.2 K. The high magnetocaloric value can be attributed to the first order phase transition of our CFO. In fact, the latent heat released/absorbed during first order conversion mainly improves magnetocaloric response[5], [6]. From another viewpoint, the improved value of ΔT can be related to the grain size as reported by [24], [25]. In another way, smaller particle sizes have more negligible magnetic hysteresis, resulting in less energy lost in the thermal process, and thus the electrocaloric effect is more significant.

For industrial refrigeration applications, the performance of MCE of CFO sample is evaluated using the Relative Cooling Power (RCP) defined by Eq. 4:

$$RCP = -\Delta S_M^{max} \times \delta_{FWHM} \quad \text{(Eq 4)}$$

Here $\Delta S_M^{max}$ refers to the maximum value of the magnetic entropy, and $\delta_{FWHM} = (T_{hot} - T_{cold})$ is the full width at half maximum of the curve $\Delta S_M^{max}$.

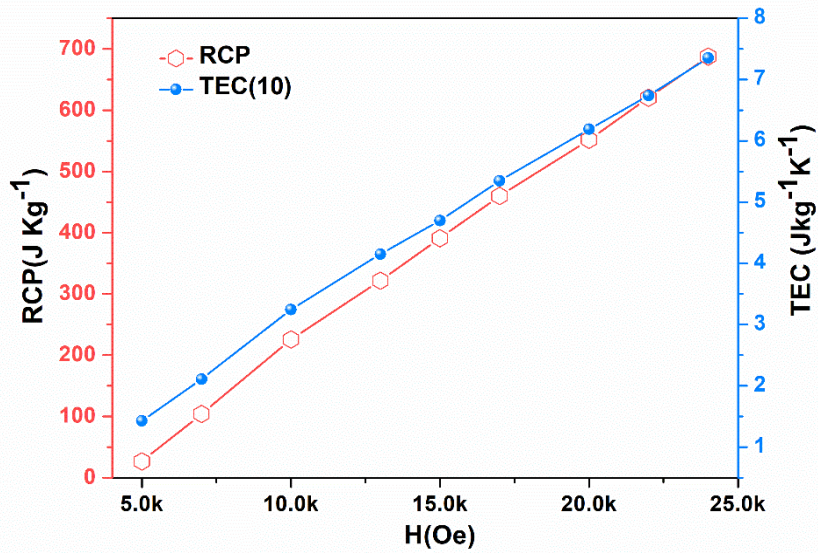

Fig. 8 Magnetic field dependence of the RCP and TEC

The RCP value increases proportionally with the applied magnetic field (Fig. 8). Therefore, the excellent RCP value of 687.56 J Kg$^{-1}$ obtained under low applied magnetic field (24 kOe) leads to a state that such a sample could be considered a potential candidate for magnetic refrigeration application at 600 K - 800 K temperature range.

To investigate the magnetic refrigeration efficiency, the ΔS, ΔT and RCP alone are insufficient to identify the potential of the refrigerant material for applications in solid-state refrigeration technologies. In this respect, the temperature-averaged entropy change (TEC) was calculated using the following expression as reported by [46]:

$$TEC(\Delta T_{H-C}) = \frac{1}{\Delta T_{H-C}} max \left\{ \int_{T_{mid}+\Delta T_{H-C}}^{T_{mid}-\Delta T_{H-C}} |\Delta S_M(T)|dT \right\}$$

Herein $\Delta T_{H-C}$ denotes the temperature difference between the device's cold and hot heat exchangers. $T_{mid}$ is the temperature at which the TEC($\Delta T_{H-C}$) is maximized for a given $\Delta T_{H-C}$. It should be mentioned that the obtained values of $\Delta S$, $\Delta T$, RCP and TEC are much higher than those reported in the literature in the ferromagnetic oxide systems (table 3). Moreover, even with a high applied magnetic field, cobalt ferrites doped with copper show a moderate $\Delta S$ value of 0.6 J Kg$^{-1}$ K$^{-1}$ with an RCP of 62.55 J Kg$^{-1}$ [41]. For Mg$_{0.6}$Cu$_{0.4}$Fe$_2$O$_4$ nanoparticles elaborated by the sol-gel technique, the maximum MCE value reported was 1.09 J kg$^{-1}$ K$^{-1}$ under an applied field of 50 kOe. In our previous work [7], a high RCP of 79.19 J Kg$^{-1}$ with a $\Delta T$ of 5.6 K under 60 kOe was achieved in perovskite manganite La$_{0.5}$Ca$_{0.5-x}\square_x$MnO$_3$. In addition, our CFO shows an important value of TEC of 7.34 J Kg$^{-1}$K$^{-1}$ compared with other magnetic materials such as La$_{0.8}$Ca$_{0.2}$MnO$_3$ ceramic (TEC =4.5 J Kg$^{-1}$K$^{-1}$) under 20 kOe, and in La$_{1.2}$Pr$_{0.2}$Ca$_{1.6}$Mn$_2$O$_7$ ceramic upon 25 kOe (TEC =6.98 J Kg$^{-1}$K$^{-1}$) [47]. This comparison makes our CFO nanopowder a good candidate for magnetic heating and cooling applications.

Table 3. Magnetocaloric properties of ferromagnetic oxides.

| Material | $T_C$(K) | Magnetic field (kOe) | $\Delta S$ (J Kg$^{-1}$ K$^{-1}$) | $\Delta T$ (K) | TEC (J Kg$^{-1}$K$^{-1}$) | RCP (J Kg$^{-1}$) | References |
|---|---|---|---|---|---|---|---|
| CoFe$_2$O$_4$ | 785 | 24 | 6.15 | 11.24 | 7.34 | 687.56 | This work |
| CoFeCuO$_4$ | 688 | 50 | 0.6 | -- | -- | 62.55 | [41] |
| Zn$_{0.2}$Ni$_{0.4}$Cu$_{0.4}$Fe$_2$O$_4$ | 705 | 50 | 1.61 | -- | -- | 233 | [8] |
| NiFe$_2$O$_4$ | 845 | 25 | 0.75 | -- | -- | 60 | [9] |
| Ni$_{0.5}$Zn$_{0.5}$Fe$_2$O$_4$ | 481 | 25 | 1.15 | -- | -- | 161 | [9] |
| Mg$_{0.6}$Cu$_{0.4}$Fe$_2$O$_4$ | 630 | 50 | 1.09 | -- | -- | 136 | [48] |
| La$_{0.8}$Ca$_{0.2}$MnO$_3$ | 105 | 20 | 0.732 | | 4.5 | 99.05 | [49] |
| La$_{0.813}$K$_{0.16}$Mn$_{0.987}$O$_3$ | 338 | 15 | 2.11 | -- | 1.47 | 77.62 | [50] |
| La$_{1.2}$Pr$_{0.2}$Ca$_{1.6}$Mn$_2$O$_7$ | 274 | 25 | 4.25 | -- | 6.98 | 110 | [47] |
| La$_{0.7}$Sr$_{0.3}$MnO$_3$ | 364 | 1 | 1.6 | -- | -- | 0.4 | [51] |
| La$_{0.5}$Ca$_{0.5-x}\square_x$MnO$_3$ | 254 | 60 | 2.70 | 5.6 | -- | 79.19 | [7] |
| Ho$_{60}$Co$_{20}$Ni$_{20}$ | 20.4 | 50 | 18.4 | -- | 18.3 | 668.2 | [52] |
| Er$_{60}$Co$_{20}$Ni$_{20}$ | 11.5 | 50 | 15.5 | -- | 15.4 | | [52] |
| EuTi$_{0.9}$Cr$_{0.1}$O$_3$ | -- | 20 | 30 | 4.2 | -- | 125 | [53] |
| EuDy$_2$O$_4$ | -17.4 | 80 | 25 | 16 | -- | 415 | [54] |

*Conclusion:*

In conclusion, we investigated the structural, magnetic, and magnetocaloric properties of $CoFe_2O_4$ ferrite (CFO) nanoparticles elaborated by sol-gel auto combustion. According to X-ray diffraction analysis and Raman spectra, the CFO crystallizes in a cubic spinel structure with a space group of $Fd\bar{3}m$. Magnetic measurements reveal that our CFO exhibits a first-order PM - FM phase transition with a Curie temperature of 785 K. The maximal magnetic entropy change and relative cooling power are both enhanced, making our CFO suitable candidate for both environmentally friendly magnetic refrigeration and medical applications that need temperatures above 300 K.